# Perspectives on the viscoelasticity and flow behavior of entangled linear and branched polymers

F. Snijkers[1,2], R. Pasquino[1,3], P. D. Olmsted[4], D. Vlassopoulos[1,5]

[1] FORTH, Institute of Electronic Structure and Laser, Heraklion, Crete, Greece
[2] CNRS/Solvay UMR 5268, Laboratoire Polymères et Matériaux Avancés, Saint-Fons, France
[3] Università degli Studi di Napoli Federico II, DICMAPI, Napoli, Italy
[4] Georgetown University, Department of Physics & Institute for Soft Matter Synthesis and Metrology, Washington DC 20057, USA
[5] University of Crete, Department of Materials Science & Technology, Heraklion, Crete, Greece

## Abstract

We briefly review the recent advances in the rheology of entangled polymers and identify emerging research trends and outstanding challenges, especially with respect to branched polymers. Emphasis is placed on the role of well-characterized model systems, as well as the synergy of synthesis-characterization, rheometry and modeling/simulations. The theoretical framework for understanding the observed linear and nonlinear rheological phenomena is the tube model, which is critically assessed in view of its successes and shortcomings, whereas alternative approaches are briefly discussed. Finally, intriguing experimental findings and controversial issues that merit consistent explanation, such as shear banding instabilities, multiple stress overshoots in transient simple shear and enhanced steady-state elongational viscosity in polymer solutions, are discussed, whereas future directions such as branch point dynamics and anisotropic monomeric friction are outlined.

## 1. Introduction

### 1.1 Overview

This article focuses on the rheological behavior of architecturally complex model polymers obtained via high-vacuum anionic synthesis [1]. Although many interesting observations have been made concerning the effects of architectural complexity on the rheological behavior of polymers, important details concerning the exact role of macromolecular structure on various rheological phenomena are still unknown. An inexhaustive list of examples includes the strong qualitative change of the linear viscoelastic response of branched polyethylenes as compared to their linear counterparts, due to long chain branching [2]; strong extensional hardening of branched polyolefins in startup uniaxial extensional flow which increased with branching content [3,4]; multiple stress overshoots in startup shear flow for highly branched styrene-butadiene rubbers [5,6]; failure of the Cox-Merz rule (comparing the dynamic oscillatory complex viscosity with steady-state shear viscosity) with the steady viscosity being above the complex one for highly branched polymers [7]; weakened shear-thinning behavior with increased branching levels in steady-state shear flow for polybutadienes [8]. Unraveling the molecular origins of these phenomena requires an in-depth knowledge of the structure-property relationships, and hence the need for well-defined macromolecular architectures, with the eventual aim to develop predictive tools.

In addition to the discussion of specific rheological effects associated with a specific macromolecular architecture, we address strong nonlinear flows in a more generic way as for certain nonlinear phenomena such as shear banding, we are only beginning to understand the behavior of the most simple linear polymers. Hence, we outline some of the outstanding issues arising in strong flows of linear entangled polymers. Despite a long history of research in this area, the combination of new advances in theoretical modeling, computer simulations, velocimetry, and new rheometric techniques has led to an increased understanding of phenomena such as shear banding, flow heterogeneities, disentanglement dynamics and convective constraint release (CCR).

The manuscript is organized as follows: we first summarize the state-of-the-art in the dynamics (linear viscoelasticity) of entangled linear and star-shaped flexible polymers. This sets the limits of our current understanding and thus prompts further questions and challenges. The following discussion will include two segments: First we discuss unresolved issues concerning the viscoelasticity of model branched polymers with more than one branch point. Then, we address the nonlinear flow of polymers having different molecular structure including simple linear polymers under one banner. This is an emerging topic where the wide variety of possible nonlinear deformations, instabilities and technical difficulties calls for the use of specific research tools and methodologies.

**1.2    Current state of entanglement dynamics**

In the past four decades we have witnessed remarkable advances in the understanding of polymer rheology and a number of excellent reviews and books are available [9-14]. The tube model is currently widely accepted as the "standard" framework for quantitatively describing the linear viscoelastic response of entangled polymers [15,16]. In a nutshell, the tube represents a mean field approximation to uncrossability constraints by neighboring chains, and the stress is assumed to be carried entirely along the backbone of the polymer chain. This ignores a viscous stress which is negligible for highly entangled chains in moderate flows, and ignores other non-bonded interactions which are important near the glass transition. For linear polymers, stress is assumed to relax via curvilinear diffusion of a single chain confined to the tube region (so-called reptation). The resulting stress relaxation modulus $G(t)$ exhibits a rubbery plateau at intermediate times and exponential decay at longer times when the chain "escapes" from its original tube on a time scale $\tau_d$, i.e. the longest relaxation time. Reptation [15] is complemented by contour-length fluctuations (CLF) [17] and mechanisms accounting for tube dilation or reorganization (originally introduced to describe polydispersity effects), such as thermal constraint release (TCR) or dynamic tube dilution (DTD) and Constraint Release Rouse (CRR) [18-22], whereas at short times longitudinal modes, due to redistribution of monomers along the tube after deformation, contribute to stress relaxation as well [21,23]. The current state-of-the-art expression, shown below, for the relaxation modulus accounts for constraint release (but not for DTD) (first term), longitudinal relaxation of segments (second term) and fast Rouse relaxation of segments within the tube (third term) [21]:

$$G(t) = G_e \left[ \frac{4}{5}\mu(t)R(t) + \frac{1}{5Z}\sum_{p=1}^{Z-1}\exp\left(-\frac{p^2 t}{\tau_R}\right) + \frac{1}{Z}\sum_{p=Z}^{N}\exp\left(-\frac{2p^2 t}{\tau_R}\right) \right] \quad (1)$$

with $G_e = \rho RT/M_e$ (with $M_e$ being the entanglement molar mass, $\rho$ the density, T the absolute temperature and R the ideal gas constant) is the entanglement (plateau) modulus, $\mu(t)$ the single chain relaxation function, R(t) the relaxation function due to constraint release, $Z=M/M_e$ the number of entanglements per chain (or tube segments), $\tau_R=\tau_e Z^2$ the Rouse time of a chain in its tube and p denotes a relaxation mode. Here, $\tau_e$ is the relaxation time of a polymer segment with length $M_e$. The software "Reptate" can easily perform fitting and calculations using the complete quantitative model and is freely available online [23]. We note for completeness that the third term describes the G(t) response at short times $t<\tau_e$. In practice the experiments do not agree well in this region because of the influence of the segmental relaxation which is often coupled to the Rouse modes, hence the moduli do not collapse and their power-law dependence on frequency is often characterized by an exponent within the range 0.5-0.7 instead of 0.5 (see also discussion of experimental results below). In the last decade most of these ingredients from the linear regime have been unified into a comprehensive microscopic model, the GLaMM model, which successfully describes a number of aspects of non-linear shear and extensional flows [24].

Branched polymers behave differently from their linear counterparts. Stars, for example (see Fig. 1a), which are the simplest branched polymers with only one single branch point, exhibit a logarithmically decaying plateau for G(t). Their relaxation can be described within the tube model as a combination of arm retraction and thermal constraint release (i.e. CLF and TCR), due to the fact that the branch point is essentially immobile [9,25,26]. The remarkable early finding was that the entire relaxation of a regular star polymer (i.e. symmetric and with a small number of arms) depends on the arm's response but not on the number of arms f (typically in the range $4 \leq f < 32$) [27]. Hence one can write the star's zero-shear viscosity as:

$$\eta_0 \approx G(\tau_{arm})\tau_{arm} \sim \left(\frac{M_{arm}}{M_e}\right)^{3/2} \exp\left(\frac{\gamma'}{2}\frac{M_{arm}}{M_e}\right) \quad (2)$$

with γ´ a constant, typically equal to 0.75 [9,12]. Note that for simplicity equation (2) represents the scaling dependence of zero-shear viscosity on the number of entanglements per arm. Detailed presentation of the full expression is beyond the scope of this review and the interested readers are referred to Refs. [9,25,26]. Macromolecules with more than one branch point, such as H-polymers, combs, star-combs, Cayley-trees and pom-pom polymers (see Fig. 1a), relax stress hierarchically with different parts relaxing at increasing time scales in a sequential fashion starting from the outermost dangling ends and moving inwards, often exhibiting a combination of TCR, CLF and eventually reptation [9,28-30], and with the relaxing segments diluting the remaining unrelaxed ones via DTD. Further details are offered in the next section. We note for completeness that recently, ring polymers, i.e., polymers without free ends, were shown not to exhibit a rubbery plateau and instead relax their stress by a power-law mechanism that cannot be described by the tube model [31,32].

A number of alternative approaches have been developed, that operate at different levels of coarse-graining. Sussman and Schweizer have developed a microscopic self-consistent theory for the dynamic transverse confinement in entangled solutions of rods without excluded volume, but including the topological constraints of entanglement. They predict a very deep potential in agreement with the tube model, however with anharmonicities [33]. The theory was extended to flexible chains by starting from a primitive path description and deriving explicit expressions for the correlated intermolecular contributions to the polymer stress that arise from chain uncrossability [34,35]. This successfully yielded a microscopic derivation of the confining tube potential, and closely reproducing the empirical relation between plateau modulus $G_e$ and packing length, $G_e \approx k_B T /p^3$ [35,36]. The packing length $p$ is defined as the ratio between polymer volume and square radius of gyration, and quantifies how a given polymer fills space [36]. We note that the tube diameter $a$ is linked to $G_e$ via the relation $G_e = k_B T n \, b^2/a^2$ with $n$ being the number of monomers per unit volume of the sample and $b$ the monomer length [37]. The theory can also account for reduced tube confinement under large deformations [34,35] and is consistent with single-molecule experiments suggesting chain localization and a distribution of tube diameters [38]. With a theory in hand that derives the tube potential, the stage is set for including the drastic correlations and changes in conformation and orientation due to non-linear flow. There are strong links between the included non-bonded stresses needed to describe the tube within this picture and the physics inherent in the glass transition

The single-mode slip-link model of Hua and Schieber [39] describes the dynamics of a chain fluctuating between constraints, the slip links, which mimic the constraints envisioned for entangled chains. This approach, which has been extended to multi-mode versions, allows for discrete events, rather than the mean-field nature of the tube, and leads to predictions comparable to the tube model [40-43]. A promising alternative is the slip-spring model of Likhtman [44] which was built on the basis of the network model of Rubinstein and Panyukov [45] and provided successful quantitative description of viscoelastic, neutron spin echo and diffusion data of different monodisperse linear polymers simultaneously. Likhtman has shown that the non-bonded stress in polymer melts can be significant [46]. As Sussman and Schweizer later discussed [34,35] this part of the stress, often thought to be isotropic and irrelevant to tube dynamics, may in fact have an important role to play in strong flows where alignment and stretching can significantly reduce the shear stress response.

Molecular dynamics simulations have helped widen our understanding of entanglements [47]. Mavrantzas, Kröger and co-workers [48,49] have employed atomistic simulations that contain all molecular details. They have been able to fully describe entanglements using recent simulation-based approaches [50,51] and, with proper mapping to tune the model, have provided accurate predictions of chain viscoelasticity for a wide range of molar masses. Zhou and Larson [52] and Bacova et al. [53] investigated the motion of branch points in asymmetric stars and other macromolecular architectures of varying chain length. Recently, Brownian dynamics (BD) simulations using the classic Kremer-Grest model [47,51] consisting of purely repulsive Lennard-Jones monomers connected by a FENE potential were proposed as an independent means to test the tube model and elucidate the origin of polymer dynamics [54].

Drastically coarse grained simulation tools were shown to be a powerful alternative to tube modeling. Briels and Padding have developed the "Twentanglements" package where coarse graining from a molecular dynamics simulation to a mesoscale results in soft repulsive potentials between blobs. An uncrossability constraint is introduced to prevent bond crossings, whereas background friction (via Langevin dynamics) or pairwise friction is used [55,56]. More recently, the "Responsive Particle Dynamics" (RaPiD) simulation method was introduced [57,58]. It considers a polymer as point particle interacting with other polymers via thermodynamic (pair potential) and transient (due to topological constraints) forces. Though the tube idea is not invoked, predictions for linear and star polymers are very close to experiments when the transient forces are appropriately tuned [59]. While lacking molecular detail, RaPiD can satisfactorily capture the wide range of timescales inherent in complex polymeric materials, both in linear and nonlinear deformations.

Closing this section we mention the so-called inverse problem or analytic rheology where one takes advantage of the extreme sensitivity of rheology to polymer architecture in order to infer the molecular structure and composition of an unknown sample. This is an ill-defined problem as the inversion of the rheological signal does not have a unique solution. Nevertheless, significant progress has been reported in the last years but its discussion is beyond the scope of this short overview article. The interested reader is referred to a recent review article [60].

## 2. Open issues in viscoelasticity of model branched polymers

### 2.1 Hierarchical Relaxation

The well-established hierarchical relaxation concept [29,30] constitutes a very powerful tool to quantify the viscoelasticity of branched polymers. Extensive comparisons of accurate experimental data obtained on model polymers with the three state-of-the-art tube-based models, i.e., the Hierarchical model (HM) [61], the Branch-on-Branch (BoB) model [62] and the Time-Marching Algorithm (TMA) [63], suggest that there are a number of important unresolved issues. Cartoon illustrations of a number of typically studied different macromolecular architectures are shown in Fig. 1a. For simplicity, we focus the rest of the discussion on model branched polymers with a comb architecture which includes more than two branch points (i.e., they depart from stars and H- or pom-pom polymers) and can mimic polymers of technological interest, such as low-density polyethylenes (LDPE) covering the range from short to long-chain branching (LCB). The rheological properties of comb polymers have been extensively discussed in the literature [64-67] and hence they serve as an excellent mindset. However, the hereafter mentioned issues are of importance for all possible branched architectures. For combs, the hierarchical relaxation picture states that, after the relaxation of the branches, the relaxed branches act as an effective solvent for the backbone (DTD), and the branch points act as sources of high friction (due to the drag associated with the relaxed branch "solvent" that is carried out by branch points during their diffusion). It is only after the relaxation of the branches, that the backbone is free to relax in the regular way (i.e. via reptation, CLF, CRR, and TCR).

The HM model was developed to describe relatively simple branched polymers such as stars, H-polymers, and combs [61,68-70]. It was the first model to consider successive relaxation of branches and backbone in a hierarchical manner, and was successfully applied to two-generation branched topologies (i.e. without branches on branches) of isoprenes and butadienes.

The BoB model is a freely available [71] computational method of handling highly polydisperse mixtures, both in terms of length distribution and branching topology [62]. A branched molecule is ordered by priority, with the outer arms relaxing first by arm retraction, like star polymers, and successive inner generations retracting inwards, until the entire molecule has relaxed. The model incorporates star-like arm retraction with tubes whose diameter increases inward with decreasing priority. Numerically, it evolves forward in time by calculating the relaxation based on the previous step. The model can take as input the predicted random molecular structures (based on reaction kinetics), and thereby successfully describe industrial-grade polydisperse branched material [72,73,74].

The TMA model [63,74,75] operates in a similar way to the BoB model, by calculating the relaxation incrementally in time, in terms of the survival probability of unrelaxed segments, in terms of relaxation times due to reptation, TCR and Rouse dynamics, which depend on the state of relaxation of the various arms and branches of the branched molecule [63,74,75]. This has been used successfully to describe simple topologies (e.g., H-polymers, pom-poms, Cayley-trees) [76] as well as polydisperse branched polyamides [74].

## 2.2 Applications to various topologies

The ingredients for the three models are the same: (i) the molecular characteristics provided by the synthesis process (molar masses and polydispersities, number of branches), (ii) rheological parameters from the experiments ($G_e$ and $M_e$, $\tau_e$), (iii) the scaling exponent for dynamic dilution ($M_e=M_e\varphi^{-\alpha}$ or $G_e=G_e\varphi^{\alpha+1}$ with $\varphi$ being the polymer volume fraction and $\alpha=1$ or $4/3$) [12,70] and the fraction $p^2$ of tube diameter for hopping of the branch point in polymers with two or more branch points (typical values for $p^2$ are 1/40 for BoB, 1/12 for HM and 1 for TMA) [9,77]. Here we note that this issue merits further investigation, and recent simulations and analysis [53,78-80] as well as selective experiments probing the branch point region (e.g., with neutron scattering) [81] have been very useful in the direction of assessing the exact role of branch point friction. The extent and validity of dynamic dilution have also been put into question and further work along these lines is currently underway [82,83]. In particular, recent investigations with linear and star polymers using a combination of rheology and dielectric spectroscopy point to a number of important details related to the role of segmental equilibration in the TCR mechanism [84,85]. For branched polymers it has only been shown that, as with polydisperse linear and star polymers, the full TCR picture does not hold [83,85]. In parallel, probe rheology has proven to be a useful test of the contribution of TCR for linear polymers in different environments [86]. Its application to combs and other model branched structures is very appealing and is currently under investigation.

Nevertheless, once these parameters are fixed, there are no further adjustable parameters in the tube model. Fig. 1b depicts the linear viscoelastic master curves of representative linear, star and comb entangled polymers of the same chemistry (polyisoprene), along with their parameter-free model descriptions (see caption for details) [87-88]. It is evident that the present predictive ability is impressive and it is also noted that, for the same chemistry, the high-frequency response is almost universal (i.e. independent of macromolecular architecture as it reflects local motion). We recall however, that in this regime the model predictions [21] deviate from the data due to the segmental relaxation (which nevertheless is similar for same chemistry and large molar masses). When the latter is appropriately subtracted, the Rouse prediction can be recovered [89].

Last but not least, polydispersity in branched polymers is not only restricted to the molar masses of the branches and backbone but extends to macromolecular structure (i.e., architecture). For example, in combs, branches can attach to the backbone at different positions and the number of branches on a certain backbone can vary from one comb to the next [91]. To this end, developments in temperature gradient interaction chromatography (TGIC) [92] have revolutionized the field and demonstrated how crucial it is to account for the presence of different architectures in a sample [93-94]. This represents a new trend in molecular rheology. The detailed quantification of the different molecular architectures in model samples not only allows for more accurate model predictions, but also helps understanding of the possible limits and conceptual problems of the tube models by assessing the discrepancies between the predictions and the experimental data [95-96], as well as better elucidating structural details of the tested polymers [97].

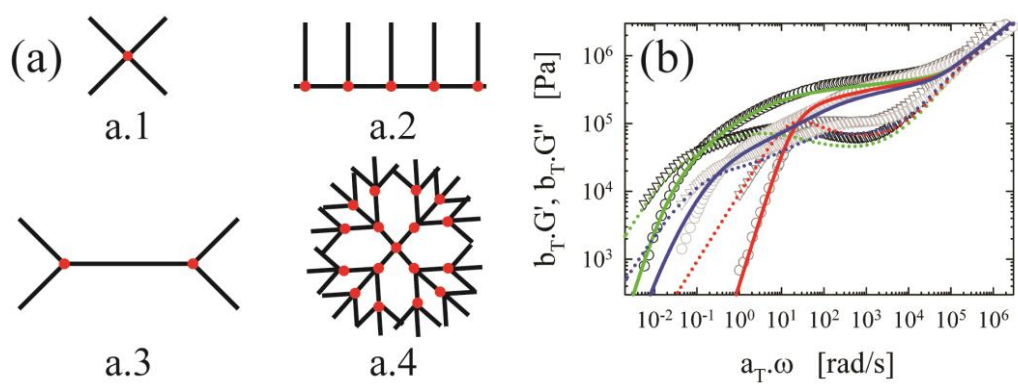

Figure 1. (a) Different complex macromolecular architectures: a.1 4-arm star; a.2 comb; a.3 H-polymer; and a.4 Cayley-tree. (b) Linear viscoelastic master curves for the elastic moduli G' (circles) and loss moduli G''(triangles) (multiplied by the vertical shift factor $b_T$) as function of angular frequency $\omega$ (normalized by the horizontal shift factor $a_T$) for three different molecular structures made of polyisoprene at T=20°C. Lines are the respective predictions using the BoB model. Linear 60kg/mol (experimental: dark gray symbols; predictions: red lines); 4-arm star with $M_a$=56kg/mol (experimental: black symbols; predictions: green lines) and comb with 5.3 arms, $M_a$=14kg/mol and $M_{bb}$=89kg/mol (experimental: light gray symbols; predictions: blue lines). Data taken from [87,88,90].

Topological polydispersity, especially the effects of different molecular architectures for branched polymers, as well as polydispersity in molecular weight, naturally brings us to the much more complex situation of commercial polymers. Recently, there have been several successful efforts to predict the viscoelasticity for several commercial branched polymers by combining knowledge from the synthesis mechanism and size-exclusion chromatography to generate a large complex mixture of different architectures that is representative for the actual sample and, via combination with one of the hierarchical tube models, accurate predictions were made for commercial LDPE's (using BoB [71,72]) and branched polyamides (using TMA [74]).

## 3. Response to strong flows

### 3.1 Challenges

The behavior of entangled polymer melts in nonlinear flow conditions remains a widely open field. Besides the practical difficulties of equilibration and sample preparation, entangled polymer melts are highly elastic and tend to become unstable in strong nonlinear deformations, which causes measurement difficulties. Instabilities include edge fracture, when strong second normal stresses can induce ejection from a free surface [98], and elastic instabilities due to inhomogeneous flows [99]. The field has evolved significantly and today it is easier to obtain reliable data for higher imposed rates of deformation or stress. In particular, for nonlinear shear deformations, the cone partitioned-plate (CPP) geometry is the optimum solution for measuring shear viscosity, and first and second normal stress differences using commercial rotational rheometers [87,100,101]. To measure strong uniaxial extensional deformations, the Filament Stretching Rheometer (FSR) [102,103] represents the state-of-the-art, whereas high-quality data can now also be obtained with commercial rheometers using the SER (Sentmanat Extensional Rheometer) fixture [104,105]. We do not discuss here the important but least developed area of controlled biaxial extension [106,107].

### 3.2 Nonlinear shear flow: Macroscopic Stress Response and Instabilities

In a simplified way, in the original tube model-approach nonlinear deformations of linear polymers were handled by three additional mechanisms: (1) orientation of tube segments in the flow direction at rates in the range $\tau_d^{-1} < \dot{\gamma} < \tau_R^{-1}$ with $\tau_d$ the longest relaxation time and $\tau_R$ the Rouse-stretch time of the chain, (2) stretch of the chain at higher rates $\dot{\gamma} > \tau_R^{-1}$ (i.e., increase of the contour-length of the chain) [108] and (3) the ability of the flow field to sweep away (or create) entanglement by their relative motion, so-called Convective Constraint Release (CCR). Doi and Edwards already recognized the first two mechanisms, while the importance of CCR was realized later [109,110]. In its simplest form CCR simply accelerates the relaxation due to reptation, according to:

$$\frac{1}{\tau_i} = \frac{1}{\tau_{i,eq}} + \beta \left( \mathbf{k} : \overline{\mathbf{S}} - \frac{1}{\lambda}\frac{d\lambda}{dt} \right) \qquad [3]$$

with $\tau_i$ the CCR-affected relaxation time, and $\tau_{i,eq}$ the original equilibrium relaxation time (i.e. unaffected by CCR), $\beta$ the CCR parameter (see below), tensor $\mathbf{k}$ the velocity gradient tensor, $S$ the average tube orientation tensor and $\lambda$ is the ratio of the stretched to unstretched tube lengths.

The effects of orientation on the nonlinear shear strain are two-fold: for shear rates exceeding the inverse of $\tau_d$ the chains are aligned in the flow direction, which leads to an eventual decreasing stress as a function of shear rate. This can lead to instabilities such as shear banding. The transient response is also expected to lead to stress maximum (or overshoot) as a function of time, or equivalently strains of order 2, as chains over-orient and hence eventually reduce the shear stress. For shear rates faster than the Rouse relaxation time the stress continues to increase as the chains stretch, hence delaying the overshoot as a function of strain. Finally CCR relaxes the chains and restores the stress, which can theoretically re-instate a stable constitutive curve without a stress maximum (see Fig. 2) [111]. This is one possible way to understand early data that did not show evidence of constitutive instability [112]. Polydispersity is expected to soften this constitutive instability because of the range of associated reptation times; recent experimental work seems to bear this out [113].

While shear banding remains controversial and ill-understood even for the "simplest" linear polymers, we are at the moment only beginning to explore the effects of molecular architecture on these different mechanisms and the macroscopic stress response. We can state that the different mechanisms (orientation, stretch and CCR) and especially the macroscopic stress response are relatively well-understood for linear, monodisperse polymers (with the currently most successful molecular theory being the GLaMM model [24]). This model uses the Likhtman-McLeish ingredients for linear response and complements them with the CCR mechanism and is able to predict the macroscopic stress response of linear monodisperse polymers in shear flow with good accuracy [114]).

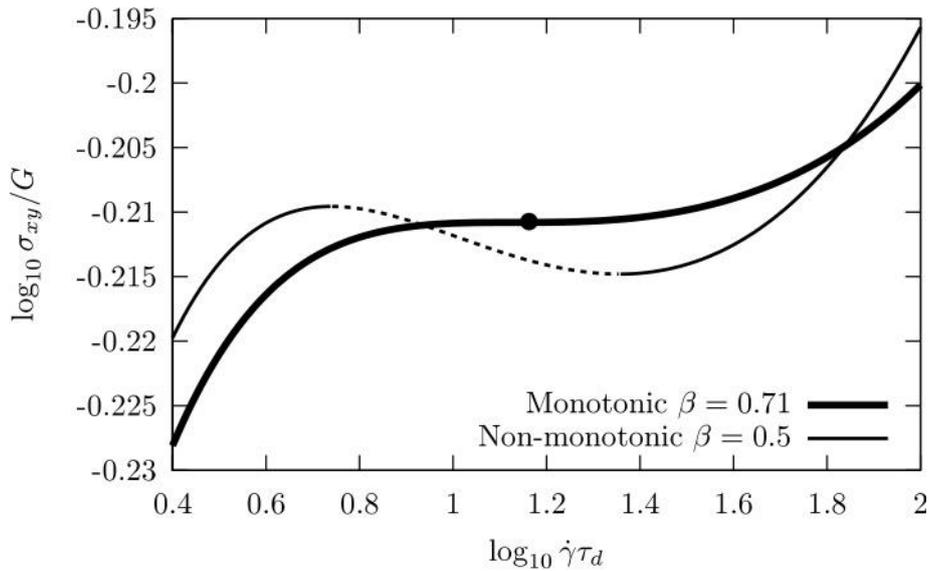

Figure 2. Example constitutive curves (shear stress as a function of shear rate) for different values of the CCR parameter β, demonstrating that sufficiently strong CCR

( $\beta = 0.71$ ) can eliminate the stress maximum associated with over-orientation of chains within the tube model (taken from [111]).

## 3.3 Rheology of Complex Topologies

Concerning the macroscopic stress response of complex model polymers, we focus our attention on the behavior in startup simple shear experiments. Fig. 3a depicts the transient response in simple shear of an entangled comb polymer with short branches, which effectively dilute the backbone in proportion to their molecular fraction; hence this is equivalent to the typical response of an entangled linear polymer [90,112,114]. On the other hand when the molar mass of branches increases for an otherwise identical comb, as in Fig. 3b, in addition to dilution the branches exhibit their own signature, which is signified by the presence of a second stress overshoot [115]. These observations are supported by predictions based on the tube model framework, and in particular invoking the branch withdrawal mechanism of the pom-pom model [116] coupled with CCR. In particular, the first overshoot at lower strain is associated with the orientation of the branches while the second one is associated with the stretch of the backbone and the stress saturation due to the gradual withdrawal of the branches into the tube of the backbone [115]. This framework explained also similar experimental evidence with commercial styrene-butadiene random copolymers [5,6,117] and the intriguing effects of rest-time on the evolution of the overshoots during repeated shear-startup tests [6]. Fig. 3 also shows another intriguing phenomenon, namely the occurrence of an undershoot between the overshoot and the steady-state stress. The undershoot has been observed systematically in the literature at high rates for both linear [114] and branched polymers [90,115] but its origin remains elusive. Also the GLaMM model predicts a slight undershoot at high rates, but much smaller than experimentally observed [24,114] and it has not been discussed. For linear polymers it is tempting to associate the undershoot with elastic recoil, but this interpretation likely does not hold for branched polymers such as combs, due to the branches and their combined role of effective solvent for the backbone entanglements and source of extra friction. In addition, normal stresses are very important as they directly reflect the elastic response of the (linear or branched) polymers. Their accurate measurement at high shear rates has been a challenge for several years. A new development of CPP using three partitions was reported recently [101] and, despite some issues for readily implementing it with polymers of interest, it constitutes the roadmap in the field.

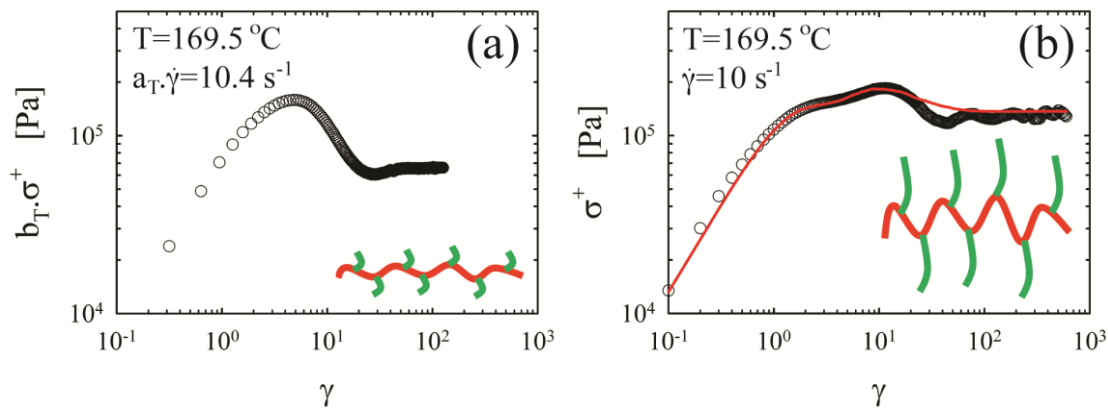

Figure 3. Start-up stress as function of strain for two different polystyrene combs at 169.5°C at a shear rate of 10 s$^{-1}$. (a) Comb PS c612 with short branches (31 arms, $M_a$=6.5kg/mol and $M_{bb}$=275kg/mol); (b) Comb PS c642 with long branches (29 arms, $M_a$=47kg/mol and $M_{bb}$=275kg/mol). The data for the PS c612 comb was obtained at 180°C and shifted to 169.5°C using the shift factors obtained from the construction of the master curves of the linear viscoelastic data. Data taken from [90,115].

The characteristic examples of Fig. 3 with combs in startup simple shear flow indicate that branching can affect the nonlinear macroscopic stress response of entangled polymers significantly (something already known for a long time [118]) and that the use of model polymers helps unraveling the detailed effects of macromolecular architecture. For example, stretch, which is controlled by the friction generated by the sliding motion of neighboring chains for linear polymers [108], can have a very different origin in branched polymers as the main part of the friction is carried by the branch points, whose motion thus controls the stretching of inner segments of the molecules [115-117]. The implications of this mechanism and the detailed effects of the molecular structure are currently of great interest. Many interesting phenomena have been observed for branched polymers, essentially due to the above-sketched idea, and these findings are now being further quantified in experiments with model polymers. In particular, the following observations have been reported for entangled comb polymers in nonlinear shear flows: failure of the Cox-Merz rule in steady-state shear flow with a steady viscosity being above the complex one for highly branched combs [119]; double overshoots in start-up shear flow at relatively low rates, as discussed above [115]; a softer damping function with increased branching levels [80,120,121]; and a large value for the third harmonic around the relaxation time of the branches in large-amplitude oscillatory shear (LAOS) [67]. Whereas some of these phenomena can be accounted for using the tube model, the precise role of molecular parameters (molar masses, number and distribution of branches, polydispersities) on each one of them is not fully decoded, so the full set of ingredients for designing macromolecules with desired performance and tunable properties remains a challenge. Another intriguing recent observation is strain hardening in startup shear for an ultra-high molar mass long-chain branched polystyrene (essentially comb polymer but with two arms grafted at each branch point) [122]. Whereas a theoretical understanding remains elusive, it was suggested that the entropic barrier to retraction mentioned above plays the key role. Clearly, this unique finding opens the route for further investigations experimental and theoretical alike. The above examples clearly illustrate that more data with well-defined branched polymers, especially having very large branching contents, in such flow conditions are desperately needed. Further studies of the effects of the rest time on repeated startup shear runs with both linear and branched polymers are also necessary [6,123-126]. Investigations on more complex deformation histories such as multiple step rate flows [127] are worth pursuing. A further, completely unexplored and somewhat elusive parameter for tailoring the flow behavior of complex macromolecules is the branch point. In particular, in highly branched polymers, such as multiarm stars, the branch point becomes non-negligible in size and represents the core on which the branches are grafted. Its chemical composition, size and shape seem to control many of the quoted phenomena [128].

### 3.4  Nonlinear Shear Flow and Shear Banding

From a different perspective, the predicted stress maximum in the Doi-Edwards constitutive curve (without sufficient CCR, as in Fig. 2) has been interpreted as a possible indication that entangled polymers will display unstable flow behavior, so-called shear-banding [129,130,131]. Incorporation of stretch within the tube does not remove the stress maximum [132], but has been proposed as a mechanism to help stabilize the flow at high shear rates [133].

Shear banding has in some cases indeed been observed, especially for polymers with a very high number of entanglements ($Z>40$) and a small molar mass between entanglements [134,135]. It is perhaps the most intriguing nonlinear phenomenon and despite the great deal of attention received with linear polymers in particular, it is still highly controversial and far from understood [111,126].

However, it is by now well-accepted that entanglements are partially swept away by the flow-field itself [136,137]. This mechanism (CCR) was historically proposed to solve the instability in the Doi-Edwards model [16,109,110,131,136,137], by preventing over-orientation of the chains in the flow direction and can lead to a stable constitutive relation, compared to the original Doi-Edwards model (Fig. 2). Recent experiments concerning the relaxation of stress from steady state for stars and combs with short branches seem to suggest that this CCR mechanism depends on architecture [88,90], and additional experiments to confirm and understand this interesting effect are needed.

Baig et al. [138] performed the first very large MD simulations of united-atom level polymers in shear flow for $Z=14$ using the p-SLODD algorithm, in which they demonstrated the loss of entanglements and the development of a bimodal distribution of chain-end separations, reflecting an excess of well-stretched chains. A modification of CCR to include the explicit dynamics entanglement loss was able to capture this effect [136]. More recent simulations of this system at higher shear rates (exceeding 50 times the retraction or Rouse relaxation rate) found a number of new timescales, including retraction and rotation of individual chains much like those of polymers in dilute solution [139]. Dissipative Particle Dynamics (DPD) simulations of a simpler model were conducted for Z=13 and Z=17, demonstrating a disentanglement transition during the stress overshoot [140]. These simulations may indicate the expected breakdown of the tube model description at high shear rates. More coarse-grained MD simulations (using the FENE bonding potential) demonstrated shear banding for moderately entangled polymer melts with Z=10 entanglements [141].

The onset of shear banding is predicted to depend on three primary quantities. For many entanglements Z, the tube model predicts a stress maximum due to over-orientation, which leads to a decreasing number of tube segments that cross the flow gradient plane and hence eventually a decreasing shear stress. Should the stress maximum occur, another mechanism for stress is needed to stabilize the system. This is posited to come from two sources: (1) the inherent solvent viscosity $\eta_s$ (or fast Rouse mode) due to intermolecular friction, which is usually negligible in comparison to stress carried by the polymer backbone, or (2) CCR, which relaxes the over-oriented chains so that they can then provide additional shear stress. Solvent viscosity (or Rouse viscosity) is parametrized by a viscosity ratio $\varepsilon = \eta_s / (G\tau_d)$, where G is the characteristic modulus, and CCR is parametrized by the CCR parameter $\beta$, which accounts for the rate at which entanglements are swept out by flow. For very small Newtonian friction (or equivalently a wide enough separation between the terminal time $\tau_d$ and the entanglement time $\tau_e$), a value $\beta$ between 0.3 and 0.55 is large enough to render the constitutive curve monotonic (for Z=10 and Z=50 entanglements, respectively) [111]. Hence, the data of [135], which show banding for Z>40 and have a solvent viscosity ratio certainly less than 0.0001, are consistent with a CCR parameter $b \gg 0.5$. This is shown in Fig. 4 for the Rolie-Poly [142] approximation to the GLaMM model. Note that even a barely stable constitutive model can give rise to rheological signatures that are consistent with shear banding [143].

Experiments similar to those of Ref. [134] were performed by Li et al. [144], for Z between roughly 40 and 90, but these authors did not observe shear banding under controlled conditions (they observed banding when there was misalignment in step strain experiments or a plastic wrapping films was used). Importantly, they controlled slippage by chemically treating the glass surfaces with crosslinked polybutadiene, with which the bulk chains could entangle to delay the onset of slip. They compared measurements in which the free surface of their cone-and-plate or parallel plate rheometer was or was not covered by a plastic wrapping film, which was the condition used by Wang et al. in their early experiments [134,145] to prevent edge fracture. Li et al. [144] found different flow profiles depending on the surface wrapping, and proposed that wrapping the outer edge could actually induce an instability from the complex flows instigated by the containing surface as had been previously suggested [98,146]. They also reported that the occurrence of the instabilities strongly depended on technical details related to the quality of the alignment of the plates. However, it was discovered later that the polymer used by Li et al. [144] had a lower molecular weight than originally quoted [147,148], while still being well-entangled. Hence, there is still an urgent need for independent studies of banding by different groups with different conditions and polymers, with and without CPP and other methods to control experimental issues. Note that wall slippage has also been proven to lead to a strain loss [144,149]. At this moment, and also due to the controversies for linear polymers, the role of controlled branching with respect to instabilities has to the best of our knowledge not been considered at all.

Very recent experiments of relaxation after large step strain have suggested that relaxation is not accelerated by chain retraction on the Rouse (retraction) timescale $\tau_R$, unlike that predicted by the GLaMM model (or other Doi-Edwards based models). If substantiated by independent results, this finding suggests that chain retraction is not free, as has been routinely assumed, but rather faces a barrier to retraction [54,150] due to topological constraints. Such a barrier could induce strain localization and an additional mechanism for shear banding. Wang and co-workers suggested that this can be the basic ingredient of an alternative theory for nonlinear rheology [151,152], similar to the approach of Sussman and Schweizer, who studied entangled rigid rod solutions undergoing nonlinear shear deformation. Sussman and Schweizer calculated that the tube field has a finite strength with an entropic barrier in the transverse direction [33,34,153]. Wang et al. further proposed that the stress overshoot in simple shear is associated with elastic yielding when the force associated with this barrier (denoted an "intermolecular gripping force") is overcome [134,154]. This "barrier" may be related to the often-neglected solvent friction (or equivalently the non-bonded stress [35,46]), which also controls the glass transition. There may be an analogy with the well-known property of small branches to have disproportionate friction-like effects in asymmetric stars [52,77]. As the microscopic origin of this yielding process remains elusive, this idea has not yet been formulated into a theory, and must still be reconciled with the clear and physical stress maximum due to over-orientation predicted by Doi and Edwards, which is "tamed" by CCR. Simulations will certainly help to further elucidate these points.

A related open issue is the interplay between chain stretch and the stress optical rule. Recent BD simulations [155] suggest that stretching occurs at Rouse-Weissenberg number $Wi_R<1$, contradicting the conclusions from the classic rheo-optical experiments of Pearson et al. [156] and multi-model slip-link simulations [157]. A challenge here is to test the experimentally observed birefringence overshoot together with lack of overshoot of orientation with BD. At the same time the definition of stretch used in BD and the comparison of radius of gyration in equilibrium and under flow may need reconsideration (among other concerns, this is not confirmed by rheo-SANS experiments [158]). We note that the number of units (Kuhn steps) in an entanglement segment under flow is not necessarily constant.

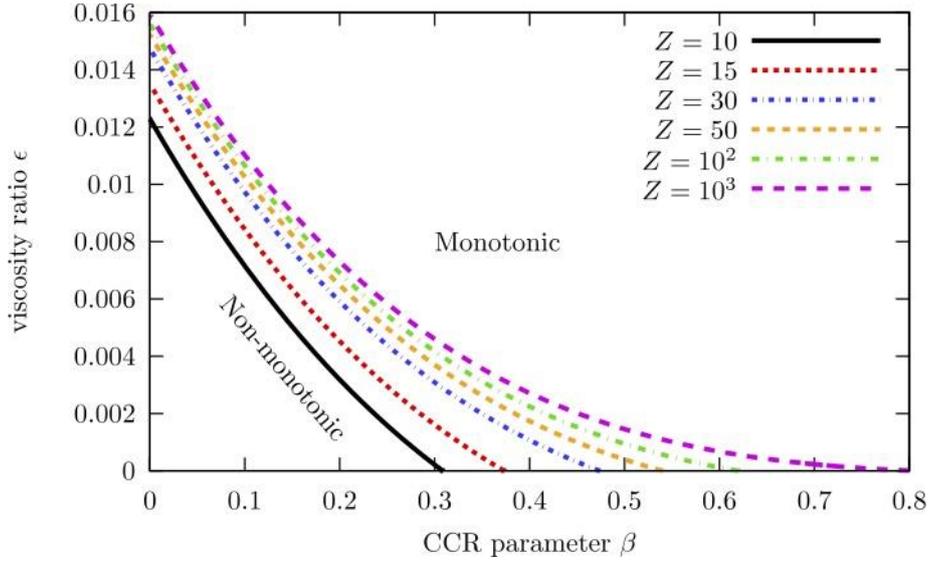

Figure 4. Regions of parameter space within which the steady state constitutive relation from the Rolie-Poly approximation to the GLaMM model is predicted to be monotonic or non-monotonic (from [111]). The data of [135] show banding for Z>40 and have a solvent viscosity ratio less than 0.0001, and are thus consistent with a CCR parameter $b \gg 0.5$.

As noted above, the tube model predicts a stress maximum as a function of time (or strain) for a given imposed shear rate exceeding the inverse terminal time $\tau_d^{-1}$. For a very rapid imposed shear rate the decreasing stress with increasing strain is equivalent to a negative differential shear modulus, which is a hallmark of an elastic instability [159], similar to that seen in the familiar necking instability of glassy polymers [160]. This should in principle give rise to instability, as was pointed out by Marrucci and Grizzuti [159], and could explain the so-called Type II anomalies seen in startup shear [160,161].

Recent particle velocimetry measurements on entangled linear polymers [134,135, 162-164] reveal that shear bands often also arise during startup, and can be sufficiently long-lived to represent the ultimate flow response of the material for practical purposes, even if the constitutive curve is monotonic. Moreover, time dependent shear banding has been reported to appear closely linked to the presence of a distinctive signature in the shape of the material's time-dependent rheological response function, and the elastic instability of [159] can be generalized to a dynamic instability closely associated with the stress overshoot [111]. Recently much effort has been devoted theoretically and numerically (using Giesekus and Rolie-Poly models) to determining precise universal criteria for transient instability beyond the simple elastic argument given above for shear banding in time dependent flows of polymeric materials [165,166].

### 3.5 Perspectives with Model Branched Polymers

Turning to model branched polymers, such as combs, the role of branches on the instabilities discussed above is unknown. Polystyrenes are the polymers of choice for achieving large shear rates without problems of edge fracture or slip [90]. These model systems come in small amounts but it appears now possible to investigate the possibly occurrence of banding instabilities in concentrated solutions. Along these lines, rheo-optical experiments [167,168] emerge as important probes of segmental orientation and the stress-optical rule whereas rheo-SANS (small angle neutron scattering) experiments using partly labeled samples will provide important insight into flow-induced deformation, as done with linear polymers. This so-called "neutron flow-mapping" provides significant insight into the nonlinear response of complex polymers over a range of length (and time) scales and, coupled with rheology and multiscale modeling, will help decoding the molecular origin of flow response in shear or extensional or mixed flows [158,169]. This is undoubtedly a central new direction in molecular rheology. In this direction, very recent single molecule studies on large DNA of varying architecture using fluorescence microscopy, elucidates their response in complex flows and can be a useful complement to respective studies of entangled polymers [170]. Extension of the abovementioned nonlinear models to branched polymers is an outstanding challenge. Very promising is also the recent RaPiD-based microscopic analysis of transient forces in star polymers and development of constitutive equations and prediction of shear banding [171].

### 3.6 Uniaxial Extensional flows: Strain Hardening, Overshoot and Steady State

Next to the macroscopic stress response of complex polymers in shear flow and the extensive and complex range of possible instabilities and their development discussed above, the understanding of the response of entangled polymers in uniaxial extension is a necessary ingredient to constitutive modeling and the eventual optimization of processing conditions. Extensional flows are in a way more interesting as polymers deform more strongly in extensional flow as compared to shear flow. Moreover the kinematics of most processing flows are extensional (fiber spinning, film blowing etc), rather than shear, in nature. As already mentioned, in recent years there has been a lot of activity in this field due to development of availability of FSR [102,103] and SER [104,105]. An important effect is the potential steady state in stress growth coefficient in uniaxial extensional flow [102,103,172]. Alvarez et al. [173], using the FSR device, demonstrated the appearance of stress maximum and eventual steady-state in extensional flow of LDPE melts.

Focusing in particular on this steady-state extensional viscosity for "simple" linear polymers, intriguing data sets concerning the behavior of polystyrene melts and concentrated solutions [102,103,174,175] triggered an interesting debate about one decade ago that continues today. While the steady extensional viscosity as function of Hencky strain rate of concentrated solutions of entangled polystyrenes followed the expectations based on the tube model, i.e. it displayed strong extensional thickening at rates above the inverse Rouse-stretch time after a region of thinning, the melts did not display the thickening region and conversely remained thinning over an extended range of extensional rates. Initially this controversy was explained with some success by the so-called "interchain tube-pressure" effect [176,177]. However, despite the success of this approach in predicting extension hardening in different polymers [178,179] the observed discrepancy between melts and solutions remained specific for polystyrenes and was e.g. not observed for other chemistries [103] and hence there is a need to search an alternative explanation for the observations. A very interesting idea has been put forward concerning the possibility of a changing monomeric friction coefficient in strong flows [42,103,179,180]. More specifically, it has been proposed that, locally, due to the strong stretch of the chains due to the flow, effects similar to nematic interactions can exist in polymers with a relatively small number of Kuhn segments between entanglements (such as polystyrene, but also e.g. polymethylmethacrylate and polyethylene) [41,103,174,175,180,181]. Consequently, the monomeric friction coefficient would reduce and become anisotropic in strong extensional flows. This hypothesis was partly confirmed by using MD-simulations on PS oligomers and slip-link simulations on several different chemistries, but it needs experimental verification and testing [179,180]. A natural question concerns the respective response of these solutions to simple shear flow at the same $Wi_R$. In fact, this is currently undergoing investigation.

Turning now to complex model samples, the strain hardening upon startup uniaxial extensional flow has been measured and quantified systematically as function of macromolecular structure for e.g. polystyrene and polyisoprene combs [67,181], with the strain hardening becoming more apparent for polymers with a larger branching content. It has been recognized that the number of entanglements of branches and segments between branches were shown to promote strain hardening for well-characterized branched polymers such as combs or Cayley trees [118,178,181,182] but, as for nonlinear shear flow, the exact link between molecular features and the response is still not fully decoded and more data on well-defined structures especially with very high branching levels is desperately needed. A modified pom-pom constitutive equation has been formulated in which dynamic dilution and branch withdrawal again provide the main physical ingredients for rationalizing the experimental data [182]. Also the behavior of complex polymers in more exotic experiments were shown to be of great interest, as e.g. in experiments concerning the relaxation upon cessation of uniaxial extensional flows [172].

A stress maximum has been reported in extensional flow of branched polymers [173], analogous to the stress maximum in entangled linear polymers in transient shear flow. In the case of long chain branched polyethylene, for example, Hoyle et al. [183] demonstrated and inferred a stress maximum followed by a steady state condition, by using both the FSR device and the cross-slot rheometer. Constitutive modeling based on the FSR data was then used to predict the W-cusp birefringence in a cross-slot device. The modeling required an empirical introduction of a stretch relaxation time that becomes longer for a greater degree of alignment, as well as a higher extension rate. The additional longer timescale accounts for the stress maximum, but still lacks firm physical foundation. The W-cusp indicates a polymeric stress shifted away from the stagnation point, which is consistent with the stress maximum in the FSR. Interestingly, the mechanism for the stress overshoot may actually be consistent with Wang's proposition of a barrier to free chain retraction [149]. For completeness, we note that Wang and co-workers reported a maximum in engineering stress, i.e., the ratio of the applied tensile force to the initial cross-sectional area (before elongation), during uniaxial extension of highly entangled linear polymer melts. For sufficiently large applied Hencky strain rate, they associated this maximum with yielding and elastic rupture, similarly to their transient shear work [184,185]. Note that this is analogous to the familiar Considère criterion [186], which describes geometric instability due to contraction, rather than constitutive instability due to molecular physics. Extension hardening is a generic phenomenon that occurs in both linear and branched polymers [187-189] as long as the Hencky strain rate exceeds the inverse stretch time (for linear polymers the stretch time is simply the Rouse time, while for branched polymers it is more complicated due to hierarchical relaxation, branch point friction and dynamic dilution [53,180,181]) [9,116,174]. Fielding's calculations [190] suggest that these experiments may be expected to go unstable via a dynamic generalization that incorporates molecular features as well as the geometric effects of the Considère criterion. This topic, along with the detailed mechanism of rupture represent one outstanding challenge is the field [191,192].

## 4. Summary and Outlook

Despite the significant developments over that last 50 years, entangled polymers remain an exciting and timely field of research. It is actually impressive that advances in the field have enhanced our understanding of polymer response to flows while at the same time a fundamental understanding of the physics of entanglements remains elusive. To this end, recent molecular dynamics simulations using concatenated or knotted rings as a vehicle for counting entanglement events (while avoiding the disentanglement process) and constructing tube trajectories, appear to hold the premise for further progress [193,194]. In conclusion, experimental and modeling advances in the outlined areas will enhance our understanding of model complex polymer melts in nonlinear deformations and this knowledge will serve to finesse the available theoretical framework and guide our understanding of commercial polymers (i.e. highly complex and very polydisperse mixtures of many architectures) to eventually obtain the much desired link of molecular design, reaction engineering, polymer processing and final properties.

Several important trends have emerged in the past years in molecular rheology. They include the detailed characterization of carefully synthesized complex polymers (often involving interaction chromatography as an indispensable tool); analysis of mixtures (for probe rheology and beyond); further tests of the limits of the current state-of-the-art tube models in nonlinear flows by combining predictions, experiments and simulations; improving/correcting the model or possibly search for alternative frameworks for nonlinear rheology; rheo-physical and single molecule experiments to gain insight into the molecular response at different scales; pushing the limits of experimentation with shear and uniaxial extension to reach higher rates and strains and explore ill-understood phenomena such are overshoots and undershoots in stress; understand in full detail the origin of transients and steady state nonlinearities due to a coupling of flow and conformation, using experiments, theory and simulations.

Moreover, driven in part by technological challenges and a desire to mimic nature, chemists have been able to create more complex structures with interesting albeit complicated rheological properties. For example, recently a lot of attention was paid to the understanding of the properties of functionalized polymers involving different interactions, hydrogen bonding being the most exploited one. The result is the formation of transient networks or supramolecular assemblies [195,196] with tunable and stimuli-responsive properties. This is becoming a huge area with immense opportunities in material design.

## 5. Acknowledgements

This overview is based on collaborative research performed in the framework of the EU-supported network MC-ITN DYNACOP (grant 214627). We are grateful to K.S. Schweizer, S.-Q. Wang and G.B. McKenna for helpful discussions.